\documentclass[namedreferences]{solarphysics}
\usepackage[optionalrh]{spr-sola-addons} % For Solar Physics
\usepackage{graphicx}
\usepackage{url}
\newcommand{\aap}{    {\it Astron. Astrophys.}}

\newcommand{\ag}{     {\it Ann. Geophys.}}

\newcommand{\apj}{    {\it Astrophys. J.}}

\newcommand{\grl}{    {\it Geophys. Res. Lett.}}

\newcommand{\jgr}{    {\it J. Geophys. Res.}}

\newcommand{\pasj}{   {\it Publ. Astron. Soc. Japan}}

\newcommand{\solphys}{{\it Solar Phys.}}

\newcommand{\ssr}{    {\it Space Sci. Rev.}}

\ifx \doiurl    \undefined \def
\doiurl#1{\href{http://dx.doi.org/#1}{\textsf{DOI}}}\fi \ifx
\adsurl \undefined \def
\adsurl#1{\href{http://adsabs.harvard.edu/abs/#1}{\textsf{ADS}}}\fi
\ifx \arxivurl  \undefined \def
\arxivurl#1{\href{http://arxiv.org/abs/#1}{\textsf{arXiv}}}\fi

\begin{document}
\begin{article}
\begin{opening}

\title{An Early Diagnostics of the Geoeffectiveness of Solar Eruptions from Photospheric
Magnetic Flux Observations: The Transition from SOHO to SDO}

\author{\inits{I.M.}\fnm{I.M.}~\lnm{Chertok}$^{1}$\orcid{0000-0002-6013-5922}} \sep
\author{\inits{V.V.}\fnm{V.V.}~\lnm{Grechnev}$^{2}$\orcid{0000-0001-5308-6336}} \sep
\author{\inits{A.A.}\fnm{A.A.}~\lnm{Abunin}$^{1}$}

  \runningauthor{Chertok, Grechnev, and Abunin}
 \runningtitle{Diagnostics of the Geoeffectiveness of Solar Eruptions}

\institute{${}^{1}$Pushkov Institute of Terrestrial Magnetism,
            Ionosphere and Radio Wave Propagation (IZMIRAN), Troitsk, Moscow, 108840 Russia,
            email: \url{ichertok@izmiran.ru}; \url{abunin@izmiran.ru}\\
            ${}^{2}$Institute of Solar-Terrestrial Physics SB RAS,
            Lermontov St.\ 126A, Irkutsk 664033, Russia, email: \url{grechnev@iszf.irk.ru}}

\date{Received ; accepted }

\begin{abstract}
In our previous articles (Chertok {\textit et al.}: 2013,
\textit{Solar Phys}. \textbf{282}, 175, and  2015, \textit{Solar
Phys}. \textbf{290}, 627), we presented a preliminary tool for the
early diagnostics of the geoeffectiveness of solar eruptions based
on the estimate of the total unsigned line-of-sight photospheric
magnetic flux in accompanying extreme ultraviolet (EUV) arcades
and dimmings. This tool was based on the analysis of eruptions
observed during 1996\,--\,2005 with the
\textit{Extreme-ultraviolet Imaging Telescope} (EIT) and the
\textit{Michelson Doppler Imager} (MDI) on board the \textit{Solar
and Heliospheric Observatory} (SOHO). Empirical relationships were
obtained to estimate the probable importance of upcoming space
weather disturbances caused by an eruption, which just occurred,
without data on the associated coronal mass ejections. In
particular, it was possible to estimate the intensity of a
non-recurrent geomagnetic storm (GMS) and Forbush decrease (FD),
as well as their onset and peak times. After 2010\,--\,2011, data
on solar eruptions are obtained with the \textit{Atmospheric
Imaging Assembly} (AIA) and the \textit{Helioseismic and Magnetic
Imager} (HMI) onboard the \textit{Solar Dynamics Observatory}
(SDO). We use relatively short intervals of overlapping
EIT\,--\,AIA and MDI\,--\,HMI detailed observations and,
additionally, a number of large eruptions over the next five years
with the 12-hour cadence EIT images to adapt the SOHO diagnostic
tool to SDO data. We show that the adopted brightness thresholds
select from the EIT 195\,\AA\ and AIA 193\,\AA\ image practically
the same areas of arcades and dimmings with a cross-calibration
factor of 3.6\,--\,5.8 (5.0\,--\,8.2) for the AIA exposure time of
2.0\,s (2.9\,s). We also find that for the same photospheric
areas, the MDI line-of-sight magnetic flux systematically exceeds
the HMI flux by a factor of 1.4. Based on these results, the
empirical diagnostic relationships obtained from SOHO data are
adjusted to SDO instruments. Examples of a post-diagnostics based
on SDO data are presented. As before, the tool is applicable to
non-recurrent GMSs and FDs caused by nearly central eruptions from
active regions, provided that the southern component of the
interplanetary magnetic field near the Earth is predominantly
negative, which is not predicted by this tool.

\end{abstract}

\end{opening}

\section{Introduction}
\label{S-introduction}

Coronal mass ejections (CMEs) and their interplanetary
counterparts, interplanetary coronal mass ejections (ICMEs), and
in particular magnetic clouds, are prime drivers of the most
severe non-recurrent space weather disturbances. Most significant
among them are major geomagnetic storms (GMSs) (\textit{e.g.}
\citealp{Gosling1993, BothmerZhukov2007, Gopal2015}) and Forbush
decreases (FDs) of the intensity of galactic cosmic rays
\citep{Cane2000, Belov2009, RichardsonCane2011}. One of the most
important challenges of solar-terrestrial physics and space
weather prediction is the diagnostics of the geoefficiency of
CMEs, \textit{i.e.} an approximate estimation and forecast of
possible non-recurrent GMS and FD parameters from observed
characteristics of an eruption that has just occurred. On the Sun,
CME eruptions are accompanied by such phenomena as bright
post-eruption arcades \citep{Kahler1977, Sterling2000,
HudsonCliver2001, Tripathi2004, Yashiro2013} and large-scale dark
dimmings \citep{Thompson1998, HudsonCliver2001, Harra2011}. They
are observed particularly in the extreme-ultraviolet (EUV) range
and represent the structures and areas involved in the CME
process.

Our previous articles (\citealp{Ch2013, Ch2015} hereafter referred
to as Article~I and Article~II) showed the total unsigned magnetic
flux of the longitudinal field at the photospheric level within
the arcade and dimming areas to be a suitable quantitative
parameter for the earliest diagnostics of the geoefficiency of
solar eruptions. This approach is based on widely accepted
concepts relating paired core dimmings to the footpoints of an
erupting CME flux rope and the post-eruption arcade to the
magnetic structures remaining after reconnection that formed this
flux rope. We studied events of Solar Cycle 23 during
1996\,--\,2005 in which sources of major non-recurrent GMSs with a
geomagnetic index Dst~$< -100$~nT were reliably identified as near
disk-center active regions (ARs). These eruptions were analyzed
using data from the \textit{Solar and Heliospheric Observatory}
(SOHO: \citealp{Domin1995}), namely solar images obtained with the
\textit{Extreme-ultraviolet Imaging Telescope} (EIT:
\citealp{Delab1995}) in the 195\,\AA\ channel and magnetograms
acquired with the \textit{Michelson Doppler Imager} (MDI:
\citealp{Scherrer1995}). As a result, clear correlations were
found between the erupted magnetic flux under the arcades and
dimmings following eruptions from ARs and the amplitude of GMSs
(Dst and Ap indexes) and FDs, as well as their temporal parameters
(intervals between the solar eruptions and the GMS onset and peak
times). The larger the erupted flux, the stronger the GMS or FD
intensities are and the shorter the ICME transit time is. These
correlations indicate that the quantitative characteristics of
major non-recurrent space weather disturbances are largely
determined by measurable parameters of solar eruptions, in
particular by the magnetic flux within the arcade and dimming
areas, and can be tentatively estimated in advance with a lead
time from one to four days.

These dependencies expressed in corresponding empirical
relationships constitute a preliminary tool based on SOHO data for
an early diagnostics of geoefficiency of solar eruptions and a
short-term forecasting of the main parameters of non-recurrent
space weather disturbances. However, at the end of 2010 July, the
synoptic 12-min cadence SOHO/EIT observations in the 195\,\AA\
channel were replaced by obtaining a couple of images per day
only, at around 01:13 and 13:13~UT (see the EIT catalog at
\url{http://umbra.nascom.nasa.gov/eit/eit-catalog.html}). In
addition, the solar magnetic field observations with the SOHO/MDI
magnetograph were terminated in 2011 April (see the MDI Daily
Magnetic Field Synoptic Data at
\url{http://soi.stanford.edu/magnetic/index5.html}).

Solar activity in the present Cycle 24 is relatively low. This has
resulted in a smaller number of non-recurrent geospace
disturbances initiated by eruptions from ARs that were relatively
weak \citep{Gopal2015}. For this reason, it is not possible to
repeat the analysis of Articles~I and II for the data of the
\textit{Solar Dynamic Observatory} (SDO: \citealp{Pesnell2012}),
which provides regular high-quality observations starting in early
2010 May. To use our tool at the present time for the early
diagnosis of solar eruptions, our procedures should be upgraded
for the usage of SDO data. To shift from SOHO data to SDO
correctly, we examine our diagnostic tool using both SOHO and SDO
observations in the intervals when they overlap. For the
extraction of the arcade and dimming areas it is reasonable to use
now the SDO \textit{Atmospheric Imaging Assembly} (AIA:
\citealp{Lemen2012}) images produced in the 193\,\AA\ channel
instead of the SOHO/EIT 195\,\AA\ images, because both EUV
channels have a close temperature response with peaks at about
1.3\,--\,1.5\,MK. Calculation of the magnetic flux in the
extracted areas should be done with data from the SDO
\textit{Helioseismic and Magnetic Imager} (HMI:
\citealp{Scherrer2012}). These issues are the subject of the
present article.

\section{Data and Methodical Issues}
 \label{S-data}

Our approach invokes a widely accepted view on the flux rope
formation in solar eruptions due to reconnection during flares.
Conjugate footpoints of an erupted flux rope are revealed by
paired core dimmings \citep{HudsonCliver2001}. Observational
studies confirm this view (see, \textit{e.g.}, \citealp{Qiu2007,
Miklenic2009}). In particular, \cite{Qiu2007} established for
several events a quantitative correspondence between the
flare-reconnected magnetic flux with the poloidal (azimuthal) flux
in the corresponding magnetic clouds near Earth. A smaller
toroidal (axial) flux corresponded to the magnetic flux in dimmed
regions.

Any erupting structure initially has magnetically conjugate bases,
even if it eventually becomes disconnected from the Sun. The
magnetic flux in one base should be the same through the erupting
structure and in the conjugated base, \textit{i.e.} the positive
and negative magnetic fluxes should be exactly balanced. A
corresponding photospheric magnetogram presents, along with the
bases of the erupting structure, numerous footpoints of compact
loops, which are not involved in the eruption. The magnetic flux
computed from a photospheric magnetogram should therefore be
basically excessive. To approach a real erupted magnetic flux,
magnetic fields should be extrapolated into the corona.
\cite{Qiu2007} extrapolated magnetic fields to a fixed height of
2\,Mm and calculated signed flux in opposite-polarity regions
separately. The flux balance defined as a ratio of the positive to
negative flux ranged typically between 0.8 and 1.5. In a case
study of \cite{Uralov2014}, the flux balance was reached by
adjusting the height of the extrapolation, and the ultimate result
corresponded well to the estimations for the near-Earth magnetic
cloud.

The last way (possibly, in an elaborated form) appears to be the
most promising for accurate estimations of the erupted magnetic
flux. However, this laborious way is difficult to use in
statistical studies such as Article~I presented. To expedite
calculations, in this article, we measure the magnetic flux
directly from photospheric magnetograms, without extrapolation.
The balanced conjugate bases of erupting structures are not known
in this situation and we use unsigned magnetic fluxes under
arcades and dimming regions for simplicity and consistency with
Articles I and II. Such measurements overestimate a real erupted
flux by a factor of two, due to the summation of the outgoing and
incoming fluxes. Our direct measurement of the magnetic flux from
the photospheric magnetograms without extrapolation causes an
additional overestimate by a factor varying from one event to
another. Thus, the actual flux in a magnetic cloud should
presumably be less than our estimates by a factor of 4\,--\,10. A
constant factor does not affect the statistical patterns we
discuss, and an unknown variable factor contributes to the
scatter, along with other circumstances, which we do not consider,
such as the sign of the southern component of the interplanetary
magnetic field, $Bz$, the presence of a possible negative $Bz$ in
the leading or trailing part of a magnetic cloud, and others.

We extract arcades and dimmings in EIT and AIA images with the
same techniques as those used in Article~I, based on formal
criteria referring to a brightness analysis, as described below.
These criteria detect the features of interest more or less
reliably; nevertheless, a manual assistance in the interactive
mode is generally required to include separate disconnected
regions or to eliminate irrelevant ones in full-disk images.

To facilitate interactive manipulations, we reduce both EIT and
AIA images by rebinning them to a common format of $512 \times
512$ pixels, as we previously handled SOHO data (Article~I). Most
EIT images and all MDI magnetograms have $1024 \times 1024$
pixels; we decrease their resolution by a factor of two.
Sometimes, EIT images are produced in a binned form of $512 \times
512$ pixels; then, these are unchanged. SDO/AIA and HMI images
have $4096 \times 4096$ pixels; their resolution is decreased by a
factor of eight. Our analysis does not require full-resolution
data, an advantage of reduced images is a slightly enhanced
signal-to-noise ratio and a smaller contribution from compact
defects.

The pixel size of the full-resolution SOHO/EIT images is
$2.63^{\prime \prime}$. SOHO is located at the L1 Lagrangian point
0.01\,AU sunward from the Earth. The EIT pixel size converted to
the near-Earth vantage point of SDO is $2.63^{\prime \prime}
\times 0.99 = 2.604^{\prime \prime}$. The ratio of the linear
scales in the EIT and AIA ($0.6^{\prime \prime}$ pixel size)
images reduced to the $512 \times 512$ pixels format is
$(2.604^{\prime \prime} \times 2)/(0.60^{\prime \prime} \times 8)
= 1.085$. We also resize MDI magnetograms to match the reduced EIT
images with a pixel size of about $5.2^{\prime \prime}$ and HMI
magnetograms to match the reduced AIA images with a pixel size of
$4.8^{\prime \prime}$. We refer henceforth to the reduced images
($512 \times 512$ pixels) with these parameters.

The data processing was carried out with IDL employing
SolarSoftware general-purpose and instrument-specific routines, as
well as a library and special software developed by the authors
for the present study. Required SOHO/EIT 195\,\AA\ and SDO/AIA
193\,\AA\ FITS files were downloaded from the NASA Solar Data
Analysis Center and Joint Science Operations Center catalogs
(\url{http://umbra.nascom.nasa.gov/ eit/eit-catalog.html} and
\url{http://jsoc2.stanford.edu/data/aia/synoptic/}). EIT images
were processed by the standard EIT\_PREP routine and we handled
level 1.5 AIA images without any additional pre-processing. We
only normalized all AIA images observed during each event to a
common exposure time, given by a pre-event image.  The solar
rotation in the analyzed images was removed and a background
pre-eruption image was subtracted from each subsequent one to
obtain fixed-base difference images.

For arcades, a criterion turned out to be appropriate, which
extracted the area around the flare site, where the brightness
exceeded 5\,\% of the maximum over the Sun during the event (in an
image with the brightest flare emission). This criterion based on
a relative brightness threshold has been used for both EIT and AIA
images without any adjustment.

The detection of dimming regions is more complex. Parameters of
dimmings were computed from the so-called ``portrait'', which
shows in a single composite image all dimmings appearing during
the event. The ``dimming portrait'' is generated by finding a
minimum brightness in each pixel over the entire fixed-base
difference set (see \citealp{ChertokGrechnev2005}).
\cite{ReinardBiesecker2008, ReinardBiesecker2009} concluded that
dimmings only appeared in events associated with fast CMEs and
strong flares. Indeed, thresholding of difference images by a
certain value (say, $-50$ for EIT images) reliably detects
dimmings in flare-related eruptions from active regions, where
pre-eruptive structures are bright, but can be insufficient in
non-flare-related events outside of active regions, where the
brightness of pre-eruptive structures is modest. To detect
dimmings in these latter events, where depressions are generally
shallower, the relative brightness thresholding is more efficient.
A brightness depression deeper than $-40\%$ of a pre-event level
is an optimal criterion for extraction of significant core
dimmings located near the eruption center and obviously related to
the eruption. On the other hand, this criterion is too strong for
some flare-related eruptions. We therefore had to use a combined
dimming-detecting criterion based on both relative and absolute
thresholds; the latter are not the same for EIT and AIA due to
their different sensitivity.

The difference in the sensitivity of the EIT and AIA telescopes
can be compensated, if a ratio is known of the AIA to EIT
responses, when both telescopes observe the same structure. Counts
per pixel in an AIA image divided by this ratio (cross-calibration
factor, CCF) should become close to those in a corresponding EIT
image, and both relative and absolute thresholds can be used for
any image independent of the instrument that produced it.

We calculated the CCF for each event as a ratio of the AIA and EIT
differences between the brightness of the quiet Sun and that of
the sky, \textit{i.e.} $\mathrm{CCF} =
(B_\mathrm{qs}-B_\mathrm{sky})_\mathrm{AIA} /
(B_\mathrm{qs}-B_\mathrm{sky})_\mathrm{EIT}$. These levels were
evaluated as the positions of the peaks of the histograms
representing the brightness distributions (number of pixels
\textit{vs.} brightness represented by the counts per pixel) in
the EIT and AIA images within a disk of $0.97 R_\odot$ (quiet Sun)
and outside $1.03 R_\odot$ (sky) from the Sun center. The peaks
are centered at the highest-probability values, around which
pixels with a highest occurrence frequency are concentrated.

The pixels corresponding to bright structures fall in the quiet
Sun histogram considerably to the right from the peak and those
corresponding to dark coronal holes fall left from the peak.
Statistical contributions of the bright and dark structures
determined by their relative areas with respect to the solar disk
are relatively small and they do not displace the peak of the
histogram. Similarly, bright off-limb structures do not affect the
position of the peak corresponding to the sky level. This reliable
technique has been widely used for calibration of microwave images
produced by radio heliographs (\textit{e.g.} \citealp{Hanaoka1994,
Grechnev2003, Kochanov2013}).

The brightness of the sky, $B_\mathrm{sky}$, is close to zero in
both EIT images pre-processed with the EIT\_PREP routine and
level 1.5 AIA images. Thus, dividing an AIA image by the
CCF brings it to the EIT data range.

\section{Comparison of Areas Involved in Eruptions}
 \label{S-areas}

Firstly, it is necessary to compare the configurations and areas
of dimmings and arcades extracted for the same events in the
SOHO/EIT 195\,\AA\ and SDO/AIA 193\,\AA\ images. Two groups of
eruptions from the central part of the solar disk (mainly within
$\pm 30^\circ$, sometimes $\pm 45^\circ$ from the disk center)
observed simultaneously with SOHO and SDO are considered
(Table~\ref{T-Table1}). Group A includes relatively weak eruptions
in 2010 May\,--\,July for which EIT images observed with a
detailed 12-min interval were still available. Because of the low
solar activity at that time, the events were selected by
inspecting daily EIT and AIA movies showing signs of an eruption
regardless of the importance and nature of the accompanying soft
X-ray flares. A number of them were classified as filament
eruptions outside ARs, although ultimately we are interested in
eruptions from ARs. In contrast, Group B consists of evident,
strong eruptions associated with flares of importance larger than
M1.0 that occurred during 2011\,--\,2015. For these events we are
forced to use the EIT images available with a 12-hour interval and
co-temporal AIA images. Considering lifetimes of arcades and
dimmings of several hours, an additional requirement for these
events was that the GOES peak of the corresponding X-ray flares
should occur not earlier than three hours before the EIT
observational time, which is usually at 01:13 and 13:13~UT. An
event was also required to be isolated in a sense that in the
considered 12-hour interval between two EIT images there were no
other events of a comparable intensity.

In most cases of Group~A, an interval starting short before the
eruption onset and lasting 3\,--\,4 hours afterwards was
considered, \textit{i.e.} a set of 15\,--\,20 images was analyzed.
During this interval, the main arcades and dimmings were already
fully formed. For events of Group B, the first image was
subtracted from the second one to reveal the arcades and dimmings
in the difference image. In some events of Group B, the time of
the second image was close to the time of the maximum of the
corresponding soft X-ray flare.

As it is known, the area of a post-eruption arcade increases with
time. Therefore, to avoid ambiguity, extraction of an arcade area
in events of Group A is performed in an image temporally close to
the maximum of the EUV flux from the selected area. Usually this
time is close to the peak time of a corresponding GOES soft X-ray
flare or somewhat later. It is clear that for Group B events the
extraction time was defined by the timing of the second EIT image
and most often it did not correspond to the maximum dimming and
arcade areas for the given event.

Table~\ref{T-Table1} lists eruptions of Groups A and B selected
for comparison of the arcade and dimming areas in the EIT
195\,\AA\ and AIA 193\,\AA\ images. Each event is specified by its
number with an index A or B in the first column, date in the
second column, time in the third column of an image used for the
arcade measurement for Group A and the soft X-ray flare peak time
for Group B, GOES class of the related flare in the fourth column,
and an approximate position of an eruption or flare in the fifth
column. We do not present the exposure times ($\tau_\mathrm{exp}$)
of the analyzed EIT images that are given in the headers of the
corresponding FITS files, because almost all EIT images (except
those for two events) had $\tau_\mathrm{exp} \approx 12.6$\,s. As
for the AIA pre-flare images, $\tau_\mathrm{exp} \approx 2.9$ and
2.0\,s are typical in the events from Groups A and B,
respectively. Below we will correct all AIA images for the
exposure times to these two typical $\tau_\mathrm{exp}$, although
some images of Group B events corresponding to the flare peaks
were produced with a much shorter exposure time of up to
$\tau_\mathrm{exp} \approx 0.07$\,s.

\begin{figure} % {1}
  \centerline{\includegraphics[width=0.75\textwidth]
   {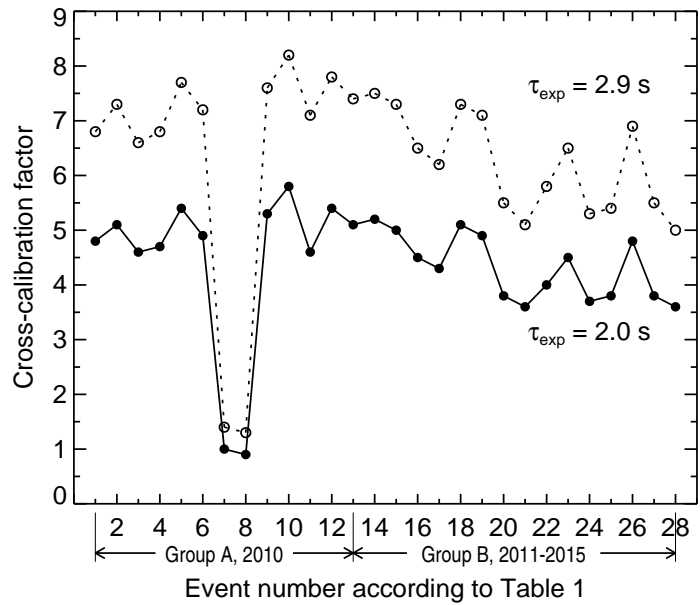}
  }
  \caption{Cross-calibration factor (CCF) between the EIT 195\,\AA\
and AIA 193\,\AA\ images with an AIA exposure time of 2.0 and
2.9\,s for Group~A and B events labeled by sequential numbers as
indicated in Table~\ref{T-Table1}.}
  \label{F-CCfactor}
  \end{figure}

The results of calculating the CCF for all events and for two AIA
exposure times typically used without flares are presented in the
sixth column of Table~\ref{T-Table1} (separated by a slash for
$\tau_\mathrm{exp} = 2.0$\,s and $\tau_\mathrm{exp} = 2.9$\,s) and
in Figure~\ref{F-CCfactor}, where the event number with an index
of A or B is specified along the horizontal axis. For AIA
$\tau_\mathrm{exp} \approx 2.0$\,s, the CCF varies from one event
to another within a span of 3.6\,--\,5.8 and for
$\tau_\mathrm{exp} \approx 2.9$\,s within a range of 5.0\,--\,8.2.
A tendency is visible to some decrease of the CCF with time on the
scale of several years; events with the smallest factor occurred
in 2014\,--\,2015. The exceptions are two Group A events, Nos. 7
and 8, observed on 2010 June 29 in which the CCF was as small as
0.9\,--\,1.4. EIT produced in these events higher-sensitivity $512
\times 512$ pixel images, while its images had $1024 \times 1024$
pixels in all other events. The CCF is expected to be directly
proportional to the AIA exposure time, \textit{i.e.}
$\mathrm{CCF}(\tau_\mathrm{exp} =
2.9)/\mathrm{CCF}(\tau_\mathrm{exp} = 2.0)$ should be equal to
$2.9/2.0 = 1.450$. The actual average ratio of the CCF for the two
exposure times is $1.440 \pm 0.027$, which characterizes the
accuracy of our cross-calibration technique.

\begin{table} %1
 \caption{Analyzed eruptions, a cross-calibration factor (CCF) between
the EIT and AIA images and areas of the extracted dimmings (Dim)
and arcades (Arc).}
 \label{T-Table1}
 \begin{tabular}{rcccccrrrr}

 \hline

No & \multicolumn{1}{c}{Date} & \multicolumn{1}{c}{Time} &
\multicolumn{1}{c}{GOES} & \multicolumn{1}{c}{Position} &
\multicolumn{1}{c}{CCF} &
\multicolumn{2}{c}{SOHO/EIT} & \multicolumn{2}{c}{SDO/AIA} \\

 & \multicolumn{1}{c}{} & \multicolumn{1}{c}{UT} & \multicolumn{1}{c}{class}  &
 &  \multicolumn{1}{c}{for $\tau_\mathrm{exp}$} & \multicolumn{4}{c}{areas [pixels]} \\

\multicolumn{5}{c}{} &  \multicolumn{1}{c}{2.0/2.9\,s} &
\multicolumn{1}{c}{Dim} & \multicolumn{1}{c}{Arc} & \multicolumn{1}{c}{Dim} & \multicolumn{1}{c}{Arc} \\

 \hline

1A  & 2010-05-23    & 19:35 & B1.3 & N18W15     &  4.8/6.8  & 1828  & 761   & 1981  & 690 \\
2A  & 2010-05-24    & 15:48 & B1.1 & N18W27     &  5.1/7.3  & 1131  & 306   & 1222  & 384 \\
3A  & 2010-05-31    & 22:24 & A6.5  & N25W27    &  4.6/6.6  & 1095  & 414   & 1303  & 363 \\
4A  & 2010-06-07    & 19:48 & B2.0  & N22E26    &  4.7/6.8  & 119   & 75    & 113   & 48 \\
5A  & 2010-06-12    & 01:14 & M2.0  & N23W43    &  5.4/7.7  & 205   & 19    & 218   & 45 \\
6A  & 2010-06-17    & 11:12 & B5.0  & N28E41    &  4.9/7.2  & 48    & 210   & 51    & 218 \\
7A  & 2010-06-29    & 13:47 & A5.0  & S20W22    &  1.0/1.4  & 413   & 375   & 625   & 477 \\
8A  & 2010-06-29    & 16:23 & B1.3  & N17W20    &  0.9/1.3  & 69    & 94    & 65    & 68 \\
9A  & 2010-07-14    & 12:48 & C1.4  & N21E06    &  5.3/7.6  & 0     & 419   & 0     & 325 \\
10A & 2010-07-16    & 15:48 & B1.3  & S21W22    &  5.8/8.2  & 922   & 49    & 875   & 44 \\
11A & 2010-07-17    & 18:23 & C2.4  & N20W33    &  4.6/7.1  & 1496  & 97    & 1419  & 109 \\
12A & 2010-07-18    & 07:12 & B1.0  & N30E20    &  5.4/7.8  & 206   & 147   & 242   & 336 \\
13A & 2010-07-19    & 09:23 & B4.0  & N30W05    &  5.1/7.4  & 1316  & 267   & 1332  & 262 \\
14B & 2011-09-06    & 22:20 & X2.1  & N14W18    &  5.2/7.5  & 1395  & 276   & 1140  & 289 \\
15B & 2012-06-13    & 13:17 & M1.2  & S16E18    &  5.0/7.3  & 1054  & 253   & 984   & 203 \\
16B & 2012-08-11    & 12:20 & M1.0  & S28W38    &  4.5/6.5  & 544   & 487   & 639   & 451 \\
17B & 2013-08-12    & 10:41 & M1.5  & S17E19    &  4.3/6.2  & 249   & 118   & 242   & 189 \\
18B & 2013-10-13    & 00:43 & M1.7  & S22E17    &  5.1/7.3  & 1126  & 111   & 1170  & 79 \\
19B & 2013-10-24    & 00:30 & M9.3  & S10E08    &  4.9/7.1  & 1783  & 80    & 1820  & 62 \\
20B & 2014-01-07    & 10:13 & M7.2  & S13E11    &  3.8/5.5  & 0     & 765   & 0     & 821 \\
21B & 2014-02-04    & 01:23 & M3.8  & N09W13    &  3.6/5.1  & 50    & 116   & 54    & 79 \\
22B & 2014-04-18    & 13:03 & M7.3  & S18W37    &  4.0/5.8  & 303   & 241   & 600   & 177 \\
23B & 2014-10-24    & 21:41 & X3.1  & S16W21    &  4.5/6.5  & 0     & 733   & 45    & 742 \\
24B & 2015-03-11    & 00:02 & M2.9  & S16E28    &  3.7/5.3  & 321   & 105   & 302   & 124 \\
25B & 2015-03-15    & 23:22 & M1.2  & S17W35    &  3.8/5.4  & 273   & 402   & 306   & 336 \\
26B & 2015-03-16    & 10:58 & M1.6  & S17W39    &  4.8/6.9  & 134   & 468   & 149   & 564 \\
27B & 2015-06-21    & 01:42 & M2.0  & N12E13    &  3.8/5.5  & 238   & 342   & 263   & 242 \\
28B & 2015-11-09    & 13:12 & M3.9  & S11E41    &  3.6/5.0  & 137   & 131   & 321   & 83 \\

 \hline

 \end{tabular}

 \end{table}

For each event we extracted the arcades and dimmings in the images
produced with both EIT and AIA and compared their characteristics,
including their quantitative areas. In the case of the AIA images,
we made it for two combinations of the exposure times (2.0 and
2.9\,s) and corresponding CCF. As noted in Section~\ref{S-data},
in the images reduced to the same common format of $512 \times
512$ pixels, the AIA pixel size exceeds the corresponding EIT
parameter by a factor of 1.085. In this study the dimming and
arcade areas are expressed in pixels. Consequently, to bring the
AIA areas to the EIT scale, the AIA areas were divided by a factor
$(1.085)^2\approx 1.18$.

\begin{figure} % {2}
  \centerline{\includegraphics[width=0.75\textwidth]
   {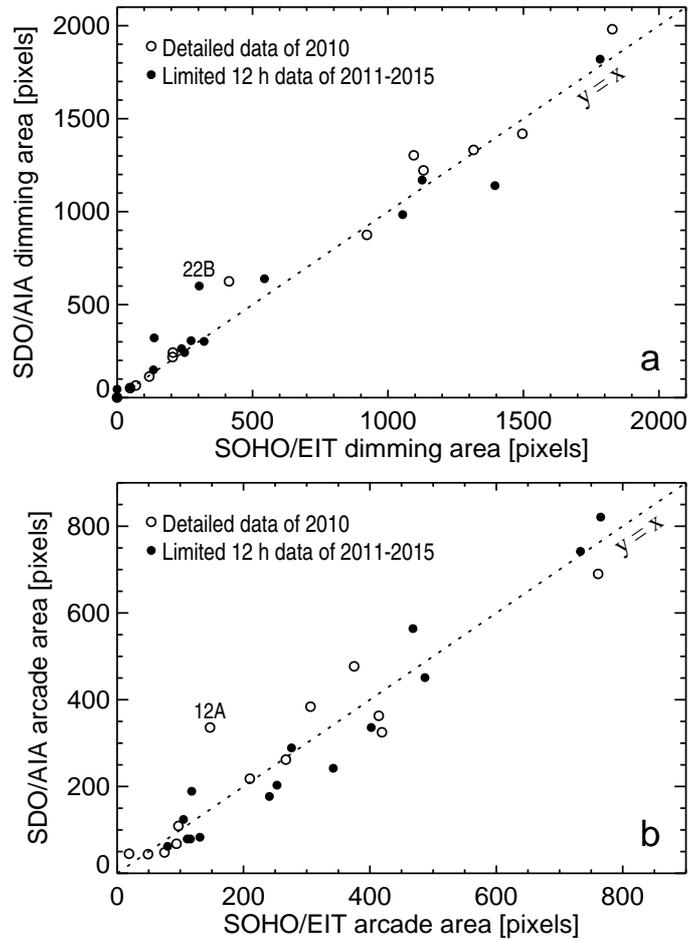}
  }
  \caption{(a) Scatter plots of dimming and (b) arcade areas
extracted in the EIT 195\,\AA\ and AIA 193\,\AA\ images. The AIA
areas are brought to the EIT pixel scale. Open and filled circles
denote events of Groups A and B, respectively. The dotted lines
correspond to the equal areas ($y = x$).}
  \label{F-areas}
  \end{figure}

One of the main results is that the areas of the arcades and
dimmings calculated from the AIA data practically do not change
with these two combinations of exposure times and CCF. In the
ninth and tenth columns of Table~\ref{T-Table1}, the AIA dimming
and arcade areas brought to the EIT pixel scale and those
extracted from the EIT images (seventh and eighth columns) are
presented. The corresponding scatter plots are shown in
Figure~\ref{F-areas}. The majority of points corresponding to the
EIT and AIA dimmings (Figure~\ref{F-areas}a) and arcades
(Figure~\ref{F-areas}b) lie near the dotted bisector ($y = x$)
line, which also nearly coincides with the best linear fit with a
correlation coefficient of 0.98 for the dimmings and 0.95 for the
arcades. We pay special attention to two events with a relatively
large difference between the EIT and AIA dimming areas (event 22B)
and arcade area (event 12A). Event 12A occurred almost without any
enhancement in the soft X-ray flux, which stayed at the B1
background level. Accordingly, its arcade was very faint and,
based on the accepted criteria, was extracted as small separate
fragments. In event 22B, the scarcely extracted weak dimmings had
a similar fragmentary character.

\begin{figure} % {3}
  \centerline{\includegraphics[width=\textwidth]
   {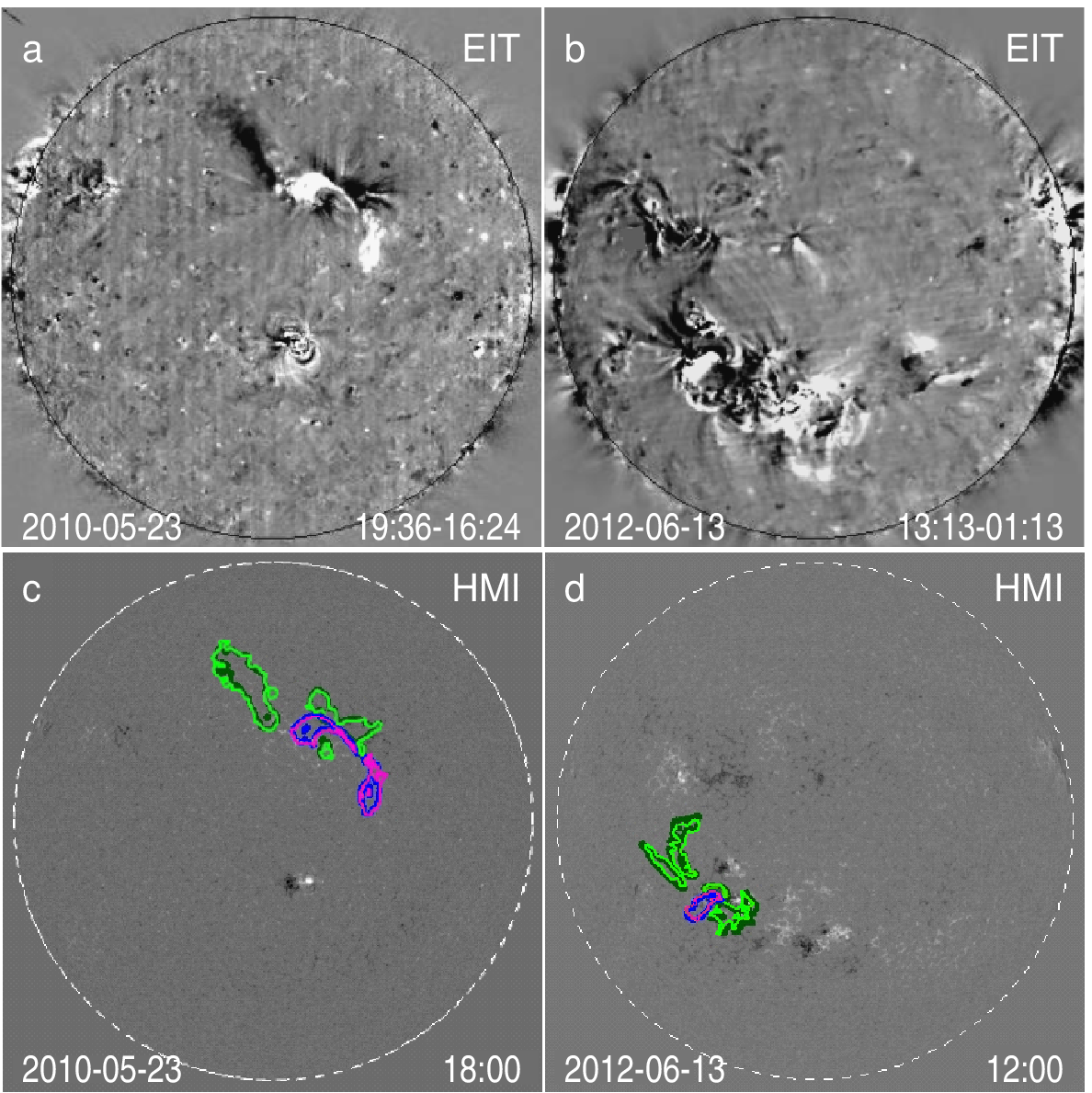}
  }
  \caption{Top: bright post-eruption arcades and dark dimmings
in SOHO/EIT 195\,\AA\ difference images for eruptions 1A (2010 May
23, panel a) and 15B (2012 June 13, panel b). Bottom:
corresponding SDO/HMI pre-event magnetograms (panels c and d)
overlaid with contours of extracted dimmings (dark green for AIA
and light green for EIT) and arcades (blue for AIA and pink for
EIT). The contours illustrate the acceptable correspondence
between the areas selected from the images produced by the two
instruments.}
  \label{F-contours}
  \end{figure}

Another important result is that the extracted arcades and
dimmings in practically all analyzed events coincide in the EIT
and AIA images both in their areas and configurations. This is
illustrated in Figure~\ref{F-contours}, where the close contours
of the EIT and AIA dimmings and arcades of two Group A eruptions
are presented by different colors on the background of the SDO/HMI
pre-event magnetograms (bottom). The AIA difference images are
very similar to those shown in Figures \ref{F-contours}a and
\ref{F-contours}b.

All of these comparisons provide a basis for a general conclusion
that with the adopted criteria, our procedures extract in the
SDO/AIA 193\,\AA\ images the same dimmings and arcades as in the
SOHO/EIT 195\,\AA\ images and AIA data can be used instead of EIT.

\section{Comparison of Erupted Magnetic Fluxes}

The second procedure, which should be considered, is the
comparison of the erupted magnetic fluxes as measured by the
SOHO/MDI and SDO/HMI magnetographs. In the period between the
beginning of HMI observations in 2010 May and end of MDI
observations in 2011 April, there were no large eruptions in ARs.
Moreover, many of the Group A events considered in the preceding
section were associated with filament eruptions outside ARs and
had very small magnetic fluxes. It is not reasonable to compare
the MDI and HMI fluxes of these eruptions. Instead, we will
compare the total magnetic fluxes of the 34 largest ARs, which
were observed near the solar disk center during the concurrent MDI
and HMI observations and had the sunspot area greater than 100
millionths of the Sun visible hemisphere ($\mu$Hem) (see,
\textit{e.g.}, the Solar Monitor site
\url{https://www.solarmonitor.org}). Although the measurements
with HMI and MDI have been compared previously, \textit{e.g.} by
\cite{Liu2012}, we are not aware of such comparisons for magnetic
fluxes erupted from large, developed active regions, where
magnetic fields are strong and saturation-like distortions are
probable.

The observation date, time, NOAA number, coordinates, and maximum
areas of these ARs are listed in Table~\ref{T-Table2}. For each of
these ARs, using data from both magnetographs, we calculated the
total unsigned magnetic flux from the line-of-sight photospheric
field within a single topologically-connected area, where the
magnetic field strength exceeded 10\,\% of the maximum value over
the entire magnetogram, but was not less than 15~G. All full-disk
magnetograms were reduced to the $512 \times 512$ pixels format.
For MDI we used level 1.8 magnetograms recalibrated in 2008
December. The corresponding MDI and HMI FITS files were downloaded
from the Stanford University sites of the MDI Daily Magnetic Field
Synoptic Data (\url{http://soi.stanford.edu/magnetic/index5.html})
and from the Stanford Joint Science Operations Center
(\url{http://jsoc2.stanford.edu/data/hmi/fits/}).

 \begin{table} %2
 \caption{The SOHO/MDI and SDO/HMI magnetic fluxes of the largest ARs over the
period from 2010 May to 2011 April.} \label{T-Table2}
 \begin{tabular}{ccccrrr}

 \hline

Date & \multicolumn{1}{c}{Time} & \multicolumn{1}{c}{AR} & \multicolumn{1}{c}{Position} & \multicolumn{1}{c}{Area} & \multicolumn{2}{c}{Magnetic flux}  \\

 \multicolumn{1}{c}{} & \multicolumn{1}{c}{hh, UT} & \multicolumn{1}{c}{number}  &
 \multicolumn{1}{c}{} & \multicolumn{1}{c}{[$\mu$Hem]} & \multicolumn{2}{c}{[$10^{20}$ Mx]} \\

\multicolumn{5}{c}{} & \multicolumn{1}{c}{$\Phi_\mathrm{MDI}$} & \multicolumn{1}{c}{$\Phi_\mathrm{HMI}$}  \\

 \hline

2010-05-24  & 08    & 11072 & S15W23    & 130   & 94    & 63 \\
2010-06-04  & 08    & 11076 & S27W42    & 190   & 72    & 51 \\
2010-07-04  & 08    & 11084 & S19W33    & 150   & 35    & 27 \\
2010-07-13  & 08    & 11087 & N12E16    & 130   & 256   & 169 \\
2010-07-22  & 08    & 11089 & S24E32    & 310   & 144   & 114 \\
2010-08-02  & 00    & 11092 & N13E07    & 290   & 109   & 70 \\
2010-08-07  & 08    & 11093 & N12E31    & 180   & 72    & 54 \\
2010-08-30  & 08    & 11101 & N12W07    & 140   & 66    & 46 \\
2010-09-17  & 08    & 11106 & S20W09    & 110   & 229   & 161 \\
2010-09-20  & 08    & 11108 & S30E22    & 420   & 142   & 115 \\
2010-09-28  & 08    & 11109 & N22W12    & 420   & 248   & 173 \\
2010-10-17  & 08    & 11112 & S18W42    & 120   & 120   & 85 \\
2010-10-18  & 08    & 11113 & N18E09    & 160   & 47    & 34 \\
2010-10-20  & 08    & 11115 & S29W01    & 190   & 48    & 30 \\
2010-10-28  & 00    & 11117 & N22W42    & 450   & 237   & 165 \\
2010-11-02  & 08    & 11120 & N39E28    & 120   & 51    & 37 \\
2010-11-16  & 08    & 11124 & N14W44    & 260   & 137   & 101 \\
2010-12-01  & 08    & 11130 & N12W41    & 190   & 130   & 95 \\
2010-12-08  & 08    & 11131 & N31W11    & 430   & 166   & 126 \\
2010-12-09  & 08    & 11133 & N14E04    & 120   & 160   & 130 \\
2011-01-02  & 08    & 11141 & N35W38    & 100   & 39    & 28 \\
2011-01-05  & 08    & 11140 & N34E01    & 210   & 108   & 55 \\
2011-02-16  & 08    & 11158 & S21W41    & 600   & 267   & 187 \\
2011-02-19  & 08    & 11162 & N18W20    & 260   & 299   & 209 \\
2011-02-21  & 08    & 11161 & N11W42    & 260   & 199   & 137 \\
2011-02-27  & 08    & 11163 & N18E32    & 110   & 59    & 28 \\
2011-03-05  & 08    & 11164 & N25W33    & 570   & 367   & 267 \\
2011-03-11  & 16    & 11166 & N09W40    & 750   & 344   & 247 \\
2011-03-13  & 08    & 11169 & N19W36    & 260   & 196   & 135 \\
2011-03-24  & 16    & 11176 & S15E44    & 490   & 200   & 155 \\
2011-03-27  & 08    & 11178 & S15E29    & 130   & 50    & 38 \\
2011-03-31  & 11    & 11183 & N15E13    & 330   & 194   & 135 \\
2011-04-05  & 16    & 11184 & N17W27    & 170   & 138   & 95 \\
2011-04-06  & 16    & 11185 & N23E36    & 100   & 37    & 28 \\

 \hline
  \end{tabular}
  \end{table}

\begin{figure} % {4}
  \centerline{\includegraphics[width=0.75\textwidth]
   {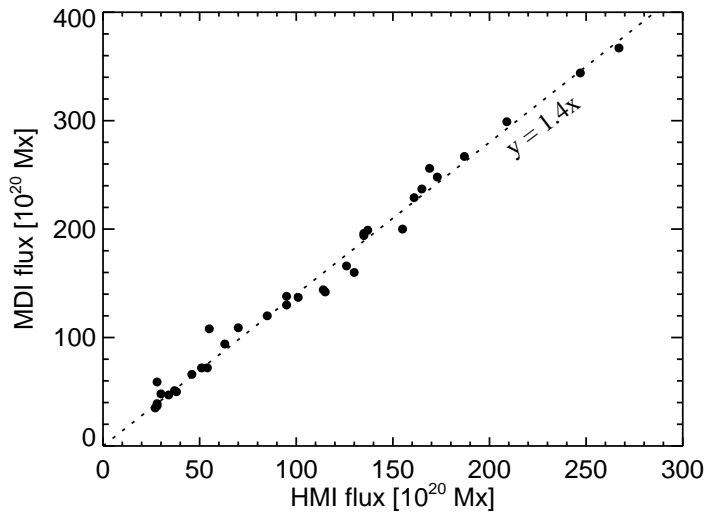}
  }
  \caption{Relationship between the MDI and HMI magnetic fluxes of the
largest ARs over the period from 2010 May to 2011 April (see
Table~\ref{T-Table2}). The dotted line corresponds to
$\Phi_\mathrm{MDI}= 1.4 \Phi_\mathrm{HMI}$.}
  \label{F-magflux}
  \end{figure}

The relationship between the MDI, $\Phi_\mathrm{MDI}$, and HMI,
$\Phi_\mathrm{HMI}$, AR magnetic fluxes calculated in this way is
shown in Figure~\ref{F-magflux}. Its best linear fit in a wide
range of the flux values of $\Phi_\mathrm{MDI} \simeq 35-350$ and
$\Phi_\mathrm{HMI} \simeq 30-270$ (in $10^{20}$~Mx units) is
$\Phi_\mathrm{MDI} = (3.1 \pm 3.5) + (1.38 \pm
0.03)\Phi_\mathrm{HMI}$ with a correlation coefficient of $r
\approx 0.99$. Within the measurement errors, the factor we
obtained from the analysis of the magnetic flux of ARs is
consistent with the result of \cite{Liu2012}, who established by a
pixel-by-pixel comparison that the line-of-sight pixel-averaged
magnetic signal inferred from MDI magnetograms was greater than
that derived from the HMI data by the same scaling factor of 1.4.
A close factor of 1.35 was also found by \cite{Svalgaard2016}. The
differences between the measurements from MDI and HMI data can be
due to their different calibration and a number of other factors
(see \citealp{Riley2014, Watson2014, Couvidat2016}). Thus, in the
transition from SOHO to SDO data, the relation $\Phi_\mathrm{MDI}=
1.4 \Phi_\mathrm{HMI}$ should be used.

\section{Transition Procedure}

The results of the two previous sections allow us to adapt the
SOHO diagnostic tool presented in Articles~I and II to SDO data
and current measurements. The updated tool does not need SOHO
data. For the extraction of dimming and arcade areas for a
particular eruption, it is sufficient to apply the
cross-calibration factor between the EIT 195\,\AA\ and AIA
193\,\AA\ images.
% In order to calculate the CCF, one of the 12-hour cadence EIT FITS
% files, the nearest to the onset of an analyzed eruption can be
% used and a corresponding pre-eruption AIA file with an exposure
% time of 2.9 or 2.0\,s. At present, two EIT files per day are not
% always available; sometimes there is one file only and sometimes
% none. In the absence of a suitable EIT file,
It is possible to simply adopt CCF~$\approx 5$ for
$\tau_\mathrm{exp} \approx 2.0$ or CCF~$\approx 7$ for
$\tau_\mathrm{exp} \approx 2.9$, based on Figure~\ref{F-CCfactor}
for the first years. After that, one should download a FITS file
of the HMI magnetogram that precedes the onset of an eruption and
a number of AIA files covering approximately the duration of an
associated soft X-ray flare. By analogy with the SOHO diagnosis,
for the extraction of the SDO dimmings and arcades it is
sufficient to take the AIA files with a 12-min interval. All AIA
images should be corrected to a single pre-event exposure time
(2.0 or 2.9\,s) and divided by the CCF before the extraction of
the arcade and dimming areas. Then, by coaligning the extracted
areas with the HMI magnetogram, the corresponding erupted flux,
$\Phi_\mathrm{HMI}$, is calculated.

Possible magnitudes as well as the onset and peak times of a
geomagnetic storm and Forbush decrease are estimated by means of
conversion of the corresponding empirical expressions, obtained in
Articles~I and II for SOHO data, in accordance with a relation
$\Phi_\mathrm{MDI}= 1.4 \Phi_\mathrm{HMI}$. For SDO data, these
expressions are as follows (again $\Phi_\mathrm{HMI}$ is in units
of $10^{20}$ Mx):

\begin{itemize}

\item
 GMS intensity (Dst and Ap indexes)
 \begin{eqnarray}
\mathrm{Dst\ [nT]} =30-15.4(\Phi_\mathrm{HMI} +3.8)^{1/2}, \nonumber \\
\mathrm{Ap\ [2nT]} = 1.12\Phi_\mathrm{HMI}. \nonumber
 \end{eqnarray}

\item
 FD magnitude
 \begin{eqnarray}
 A_\mathrm{F}\ [\%] = -0.3 + 0.042\Phi_\mathrm{HMI}. \nonumber
 \end{eqnarray}

\item The onset ($\Delta T_\mathrm{o}$) and peak ($\Delta
T_\mathrm{p}$) transit times, \textit{i.e.} the intervals between
the eruption time (maximum time of an associated soft X-ray burst)
and the start and peak of a corresponding GMS
 \begin{eqnarray}
 \Delta T_\mathrm{o}\ [\mathrm{h}] = 98/(1 + 0.00616\Phi_\mathrm{HMI}),  \nonumber\\
 \Delta T_\mathrm{p}\ [\mathrm{h}] = 118/(1 + 0.0056\Phi_\mathrm{HMI}).  \nonumber
 \end{eqnarray}

\end{itemize}

\section{Examples}

Now we consider some examples of an SDO-based post-diagnosis
related to several large eruptions in ARs located not far from the
solar disk center and to major GMSs, which occurred during the
current Solar Cycle 24. As already noted, due to a relatively low
level of solar activity after 2009, only few major GMSs with
Dst~$< -100$~nT occurred. Most of the GMSs were caused by filament
eruptions outside ARs and sometimes by high-speed solar wind from
coronal holes \citep{Gopal_et_al_2015, Gopal2015}. For instance,
the strongest geomagnetic storm of the current solar cycle with a
minimum Dst~$= -228$~nT on 2015 March 17 was initiated by a large
south-west filament eruption near AR~12297. Note that, according
to our estimations, its erupted magnetic flux,
$\Phi_\mathrm{HMI}\approx 90.5 \times 10^{20}$~Mx (corresponding
to the flux measured by MDI of $\Phi_\mathrm{MDI}\approx 126.7
\times 10^{20}$~Mx), was larger than the fluxes in filament
eruptions, which caused the GMS during Solar Cycle 23 (blue
triangles in Figure~4 of Article~I).

Table~\ref{T-Table3} presents the results of the SDO-based
diagnosis of five flare-associated eruptions in ARs carried out
according to the procedure described in the previous section. The
table lists parameters of a solar event in the second to sixth
columns, including the date, peak time, coordinates, and GOES
class of a related flare (second to fifth columns), and a total
unsigned magnetic flux calculated from HMI magnetograms in
dimmings and arcades in the sixth column. The seventh to
nineteenth columns are related to the geospace disturbance,
including its peak date and time, and estimated (letter code
``est'') and observed (``obs'') parameters of a corresponding GMS
and FD. Information on the observed hourly Dst index was taken
from the WDC2 Kyoto service
(\url{http://wdc.kugi.kyoto-u.ac.jp/dstdir/index.html}), while the
values of the three-hour Ap index were estimated in the
GeoForschungsZentrum (GFZ), Potsdam
(\url{ftp://ftp.gfz-potsdam.de/pub/home/obs/kp-ap/wdc/}). The
onset transit time, $\Delta T_\mathrm{o}$, is defined as an
arrival time of the corresponding interplanetary disturbance
(shock wave) at Earth indicated by the geomagnetic storm sudden
commencement (SSC) (\url{http://www.obsebre.es/en/rapid}).
Following Article~I, the FD maximum magnitude is adopted; this
magnitude corresponds to a cosmic ray rigidity of 10~GV evaluated
from data of the world network of neutron monitors using the
global survey method (\citealp{Krymskii1981, Belov2005}).
Additionally, the eighteenth and nineteenth columns  of
Table~\ref{T-Table3} list the hourly strength of the total
interplanetary magnetic field near the Earth, $Bt$, and its $Bz$
component according to the Operating Missions as a Node on the
Internet (OMNI) data
(\url{ftp://spdf.gsfc.nasa.gov/pub/data/omni/low_res_omni/}).

\begin{kaprotate}

 \begin{table*}    %3
 \caption{Results of the SDO-based post-diagnosis of some eruptions of Solar Cycle 24.}
\label{T-Table3}
 \begin{tabular*}{\maxfloatwidth}{cccccrcrrrrrrrrrrrr}

 \hline

No & \multicolumn{5}{c}{Eruption} & \multicolumn{9}{c}{Geomagnetic storm} & \multicolumn{2}{c}{Forbush} & \multicolumn{2}{c}{Magnetic} \\

  & \multicolumn{5}{c}{} & \multicolumn{9}{c}{} & \multicolumn{2}{c}{decrease} & \multicolumn{2}{c}{field [nT]} \\

  & \multicolumn{1}{c}{Date} & \multicolumn{1}{c}{Time} &
\multicolumn{1}{c}{Position} & \multicolumn{1}{c}{Flare} &
 \multicolumn{1}{c}{$\Phi_\mathrm{HMI}$} & \multicolumn{1}{c}{Peak} & \multicolumn{2}{c}{Dst [nT]} &
 \multicolumn{2}{c}{Ap [2nT]} & \multicolumn{2}{c}{$\Delta T_\mathrm{o}$ [h]} & \multicolumn{2}{c}{$\Delta T_\mathrm{p}$ [h]} & \multicolumn{2}{c}{[\%]} &
 \multicolumn{1}{c}{$Bt$} & \multicolumn{1}{c}{$Bz$} \\

 & \multicolumn{1}{c}{} & \multicolumn{1}{c}{UT} &
\multicolumn{1}{c}{} & \multicolumn{1}{c}{class} &
\multicolumn{1}{c}{[$10^{20}$ Mx]} & \multicolumn{1}{c}{dd/hh} &
\multicolumn{1}{c}{est} & \multicolumn{1}{c}{obs} &
\multicolumn{1}{c}{est} & \multicolumn{1}{c}{obs} &
\multicolumn{1}{c}{est} & \multicolumn{1}{c}{obs} &
\multicolumn{1}{c}{est} & \multicolumn{1}{c}{obs}  &
\multicolumn{1}{c}{est} & \multicolumn{1}{c}{obs}   &
\multicolumn{1}{c}{} & \multicolumn{1}{c}{ } \\

 \hline

1 & 2012-03-07 & 00:24 & N17E27 & X5.4 & 178.5 & 09/09 & -178 & -131 & 200 & 132 & 47 & 35 & 59 & 56  & 7.2 & 11.7 & 23.1 & -16.1 \\

2 & 2012-07-12 & 16:49 & S15W01 & X1.4 & 197\ \ & 15/19 & -188 & -127 & 220 & 132 & 44 & 49 & 56 & 65  & 8.0 & 6.4 & 27.3 & -17.7 \\

 3 & 2013-03-15 & 06:58 & N11E12 & M1.1 & 85.5 & 17/21 & -116 & -132 & 96 & 111 & 64 & 47 & 80 & 62  & 3.3 & 4.6 & 17.8 & -14 \\

 4 & 2015-06-21 & 02:06 & N12E13 & M2.6 & 234.4 & 23/05 & -208 & -204 & 263 & 236 &40 &40   & 51    &52 & 9.5   & 8.4   & 37.7 &    -26.3 \\

 5 & 2015-12-28 & 12:45 & N19W22    & M1.8  & 128.3 & 01/01 & -147 & -117 & 144 & 80 & 55   & 63    & 69    & 84    & 5.1   & 4.3\tabnote{The observed FD magnitude was evaluated
 from data of two high-latitude different-hemisphere stations, Thule and McMurdo.} & 16.9 & -15.8 \\

 \hline

 \end{tabular*}

 \end{table*}

 \end{kaprotate}

Table~\ref{T-Table3} confirms the results of Articles~I and II
that the early diagnosis of the eruptions, based on the erupted
magnetic flux evaluated in this case from SDO data, provides an
approximate assessment for the importance of the related space
weather disturbances. Particularly, the estimated values of Dst,
Ap, $\Delta T_\mathrm{o}$, $\Delta T_\mathrm{p}$, and FD are
comparable with the observed ones. According to the NOAA space
weather scale
(\url{http://www.swpc.noaa.gov/noaa-scales-explanation}), four
events (Nos. 1\,--\,3 and 5) are classified as G2\,--\,G3
(moderate\,--\,strong) storms, and one (No. 4) as G4 (severe)
storm, judging from both the estimated and observed GMS intensity.

Table~\ref{T-Table3} presents an apparent scattered correspondence
between the estimated and observed parameters of the GMSs and FDs.
Indeed, the larger the erupted flux, the stronger the actual
intensity of the GMSs and FDs, and shorter the time intervals
between the parent eruption and the GMS onset and peak are. For
example, the Pearson correlation coefficient between the estimated
Dst and its observed value is 0.61. On average,
$\mathrm{Dst}_\mathrm{obs} / \mathrm{Dst}_\mathrm{est} = 0.87 \pm
0.19$ ($\pm 22\%$), while the role of the $Bz$ component seems to
be comparable, $-Bz/Bt = 0.75 \pm 0.11$ ($\pm 15\%$). Thus, the
differences between the estimated and observed Dst can be mostly
due to the unaccounted $Bz$. There is also a close correspondence
with a correlation coefficient of 0.91 between the erupted flux
(sixth column) and the total magnetic field strength ($Bt$ in the
eighteenth column) brought to Earth by the ICMEs.

The event on 2016 June 21 (No.~4 in Table~\ref{T-Table3}) with the
largest erupted flux resulted in the most intense GMS and FD,
having also the shortest transit times and strongest
interplanetary magnetic field. In some events, for example Nos. 1
and 2, the estimated GMS intensity, measured both by the Dst and
Ap indexes, markedly exceeds the observed one, while the
magnitudes of the FDs are much closer. This is apparently due to
the fact that in these cases the negative $Bz$ component, which
determines the GMS intensity, but not considered in our
preliminary tool, amounts to only part of the total interplanetary
magnetic strength, which determines the FD magnitude.

As for the temporal parameters of GMSs, a close correspondence
between the estimated and observed values of both $\Delta
T_\mathrm{o}$ and $\Delta T_\mathrm{p}$ is present for event
No.~4. In other cases, some differences can be seen. Our simple
tentative estimates have been done as if they were issued right
after an eruption, without taking into account preceding activity,
actual magnetic field and plasma distributions in the corona, ICME
drag in the solar wind, and other factors.

\section{Concluding Remarks}

We compared quantitative parameters of CME-associated
post-eruption arcades and dimmings observed with EUV telescopes
and magnetographs aboard the SOHO and SDO spacecraft. Two basic
facts have been established. First, with the adopted thresholds of
relative brightness changes, practically the same arcade and
dimming areas are extracted from EIT 195\,\AA\ and AIA 193\,\AA\
images, if their cross-calibration factor in a range of
3.6\,--\,5.8 and 5.0\,--\,8.2 is taken into account for the AIA
exposure time 2.0 and 2.9\,s, respectively. Second, for the same
photospheric areas of strong magnetic fields in large active
regions, the MDI line-of-sight magnetic flux systematically
exceeds the HMI flux by a factor of 1.4.

These results allowed us to upgrade the tool for the early
diagnostics of AR eruptions described in Articles~I and II for
SOHO data to the current SDO observations. Empirical relationships
are obtained to connect the erupted magnetic flux measured from
SDO data with possible intensity and temporal parameters of
forthcoming non-recurrent GMSs and FDs. The case studies presented
here confirm that the updated diagnostic tool based on SDO data
also produces acceptable results, providing a prompt and
sufficiently correct assessment of the importance of forthcoming
space weather disturbances using only magnetic flux within the
arcade and dimming areas. The tool presented here and previously
in Articles~I and II corroborates the idea that parameters of
solar eruptions, CMEs and ICMEs, and geospace disturbances are
largely determined not only by the characteristics of associated
ARs and flares, but by a measurable quantity as this erupted
magnetic flux in arcades and dimmings (see Article~I;
\citealp{Demoulin2008, Mandrini2009} for a review).

It is clear, however, that the proposed tool for the early
preliminary prognostic estimations does not take into account many
factors affecting the GMS and FD characteristics. Consequently, as
in Articles~I and II, we do not pursue  exact estimates of the
parameters of GMSs and FDs and focus instead on their possible
importance. In practice, our tool should be the starting point of
a complex of comprehensive forecasting tools, which would consider
information on near-the-Sun CMEs, various models of eruptions and
drag of ICMEs propagating in the solar wind, stereoscopic
observations, estimation of a probable sign of the $Bz$ component,
and others (see, \textit{e.g.}, \citealp{Gopal2015}, and
references therein).

\begin{acks}

We appreciate the painstaking work of the anonymous reviewer for
valuable remarks and recommendations that significantly helped us
to bring this article to its final form. The authors thank the
SOHO/EIT and MDI and SDO/AIA and HMI teams for their open data
used in our study. SOHO is a project of international cooperation
between ESA and NASA. SDO is a mission of the NASA's Living With a
Star (LWS) Program. We are grateful to A.V.~Belov for his
assistance and useful discussions. This research was partially
supported by the Russian Foundation of Basic Research under grant
14-02-00367.

\end{acks}

\section*{Disclosure of Potential Conflicts of Interest} The authors
claim that they have no conflicts of interest.

\end{article}

\end{document}